\documentclass[prl,twocolumn,fleqn]{revtex4}
\usepackage{amscd,verbatim}
\usepackage[all]{xy}
\usepackage{graphicx}

\usepackage{amsmath}
\usepackage[psamsfonts]{amssymb}
\usepackage{euscript}

\usepackage{latexsym}

\def\hsp{,\hspace{.7cm}}

\def\th#1#2{\ensuremath{\theta_{#1#2}}}

\def\cp#1#2#3{\hbox{\rm cos}^#1(\th#2#3)}
\def\sp#1#2#3{\hbox{\rm sin}^#1(\th#2#3)}

\def\spp#1#2#3#4{\hbox{\rm sin}^#1(#2\th#3#4)}

\def\m#1#2{\ensuremath{\Delta M_{#1#2}^2}}
\def\mn#1#2{\ensuremath{|\Delta M_{#1#2}^2}|}

\def\meff{\ensuremath{\Delta M^2_{\rm{eff}}}}
\renewcommand{\cos}{\textrm{cos}}
\renewcommand{\sin}{\textrm{sin}}

\renewcommand{\(}{\begin{equation}}
\renewcommand{\)}{end{equation} \vspace{-.05in}\linebreak}

\newcounter{saveeqn}
\newcounter{savealpheqn}

\newcommand{\alpheqn}{\setcounter{saveeqn}{\value{equation}}%
  \stepcounter{saveeqn}\setcounter{equation}{0}%
  \renewcommand{\theequation}{\mbox{\arabic{section}.\arabic{saveeqn}
\alph{equation}}}
  \renewcommand{\)}{\end{equation}}}
\def\part#1{\frac{\partial}{\partial{#1}}}%
\def\group#1{\refstepcounter{equation}\setcounter{saveeqn}
 {\value{equation}}%
  \label{#1}\setcounter{equation}{0}%
\renewcommand{\theequation}{\mbox{\arabic{section}.\arabic{saveeqn}
\alph{equation}}}
  \renewcommand{\)}{\end{equation}}}
\newcommand{\reseteqn}{\setcounter{equation}{\value{saveeqn}}%
  \renewcommand{\theequation}{\arabic{section}.\arabic{equation}}%
  \renewcommand{\)}{\end{equation}}}

\newcommand{\aalpheqn}{\setcounter{saveeqn}{\value{equation}}%
  \stepcounter{saveeqn}\setcounter{equation}{0}%
  \renewcommand{\theequation}{\mbox{
        \Alph{subsection}.\arabic{saveeqn}\alph{equation}}}
   \renewcommand{\)}{\end{equation}}}
\newcommand{\areseteqn}{\setcounter{equation}{\value{saveeqn}}%
  \renewcommand{\theequation}{\Alph{subsection}.\arabic{equation}}%
  \renewcommand{\)}{\end{equation}}}

\renewcommand{\thefootnote}{\alph{footnote}}
\renewcommand{\(}{\begin{equation}}
\renewcommand{\)}{\end{equation}}
\newcommand{\ba}{\begin{eqnarray}}
\newcommand{\ea}{\end{eqnarray}}

\newcommand{\bp}{\mathop{\vtop{\ialign{##\crcr
   $\hfil\displaystyle{}\hfil$\crcr\noalign{\kern-13pt\nointerlineskip}
   \BIG{(}\hskip0pt\crcr\noalign{\kern3pt}}}}}
\newcommand{\cbp}{\mathop{\vtop{\ialign{##\crcr
   $\hfil\displaystyle{}\hfil$\crcr\noalign{\kern-13pt\nointerlineskip}
   \BIG{)}\hskip0pt\crcr\noalign{\kern3pt}}}}}
\newcommand{\pa}{\mathop{\vtop{\ialign{##\crcr
    
$\hfil\displaystyle{\oplus}\hfil$\crcr\noalign{\kern+1pt\nointerlineskip 
}
   \hspace{.08in}$^{\alpha=0}$\hskip6pt\crcr\noalign{\kern3pt}}}}}
\renewcommand{\hsp}{,\hspace{.3in}}

\newcommand{\beq}{\begin{equation}}
\newcommand{\eeq}{\end{equation}}

\renewcommand{\theequation}{\mbox{\arabic{equation}}}

\def\hsp#1{\hspace{#1in}}

\catcode`\@=11
\def\vereq#1#2{\lower3pt\vbox{\baselineskip1.5pt \lineskip1.5pt
\ialign{$\m@th#1\hfill##\hfil$\crcr#2\crcr\sim\crcr}}}
\catcode`\@=12

\makeatletter
\newcommand\figcaption{\def\@captype{figure}\caption}
\newcommand\tabcaption{\def\@captype{table}\caption}
\makeatother
\renewcommand{\(}{\begin{equation}}
\renewcommand{\)}{\end{equation}}

\def\th#1#2{\ensuremath{\theta_{#1#2}}}

\def\cp#1#2#3{\hbox{\rm cos}^#1(\th#2#3)}
\def\sp#1#2#3{\hbox{\rm sin}^#1(\th#2#3)}

\def\spp#1#2#3#4{\hbox{\rm sin}^#1(#2\th#3#4)}

\def\m#1#2{\ensuremath{\Delta M_{#1#2}^2}}
\def\mn#1#2{\ensuremath{|\Delta M_{#1#2}^2}|}

\def\meff{\ensuremath{\Delta M^2_{\rm{eff}}}}

\renewcommand{\beq}{\begin{equation}}
\renewcommand{\eeq}{\end{equation}}
\newcommand{\bea}{\begin{eqnarray}}
\newcommand{\eea}{\end{eqnarray}}
\newcommand{\beas}{\begin{eqnarray*}}
\newcommand{\eeas}{\end{eqnarray*}}

\newcommand{\bquo}{\begin{quote}}
\newcommand{\enqu}{\end{quote}}
\def\hsp{,\hspace{.2cm}}

\newcommand{\pic}{\hspace{-.05cm},\hspace{-.05cm}}

\begin{document}
% ======================================================================== 
\def\thefootnote{\fnsymbol{footnote}}

\title{The Neutrino Mass Hierarchy from Nuclear Reactor Experiments}

\author{Emilio Ciuffoli, Jarah Evslin and Xinmin Zhang}
\affiliation{TPCSF, IHEP, 
CAS, YuQuanLu 19B, Beijing 10049, China 
}

\begin{abstract}
\noindent
10 years from now reactor neutrino experiments will attempt to determine which neutrino mass eigenstate is the most massive.   In this letter we present the results of more than seven million detailed simulations of such experiments, studying the dependence of the probability of successfully determining the mass hierarchy upon the analysis method, the neutrino mass matrix parameters, reactor flux models, geoneutrinos and, in particular, combinations of baselines.  We show that a recently reported spurious dependence of the data analysis upon the high energy tail of the reactor spectrum can be removed by using a weighted Fourier transform.  We determine the optimal baselines and corresponding detector locations.  For most values of the CP-violating, leptonic Dirac phase $\delta$, a degeneracy prevents NO$\nu$A and T2K from determining either $\delta$ or the hierarchy.  We determine the confidence with which a reactor experiment can determine the hierarchy, breaking the degeneracy.

\end{abstract}

% \vfill
% 
% \end{titlepage}
\setcounter{footnote}{0}
\renewcommand{\thefootnote}{\arabic{footnote}}

% \pacs{??}

\maketitle

% \pagebreak
% =====================================================================

\section{Introduction}

Last year the Daya Bay \cite{dayabay,neut2012} and RENO \cite{reno} experiments demonstrated that the neutrino mass matrix mixing angle $\theta_{13}$ is nonzero and several times larger than had been suspected just two years earlier.   With the discovery that  $\theta_{13}$ is nonzero, at least three qualitative questions remain to be answered in the standard model plus three massive neutrinos.  First, it is not known whether the second or third neutrino flavor is the most massive.  This choice is known as the neutrino mass hierarchy.  Second, it is not known whether the leptonic sector has a non-zero CP-violating Dirac phase $\delta$.  Third, the most precise determination of the mixing angle, $\theta_{23}$, by the MINOS collaboration is \cite{minosneut2012}
\beq
\spp2223=0.96\pm 0.04.
\eeq
This equation has eight distinct solutions, known as the octants of $\theta_{23}$.   Two of these solutions lead to inequivalent mass matrices, corresponding to $\theta_{23}$ modulo $90^\circ$  greater than or less than  $45^\circ$.

The large value of $\theta_{13}$ means that an appearance experiment, searching for $\nu_e$'s in a beam of accelerator $\nu_\mu$'s, may be sensitive to CP violation in the leptonic sector.  The trouble is that, at the baselines of the T2K experiment in Japan and the NO$\nu$A experiment in the US, the CP phase $\delta$ is degenerate with the mass hierarchy, the octant of $\theta_{23}$ and the precise value of $\theta_{13}$.   Following the analysis in Ref.~\cite{octant2013}, by combining data from the appearance of $\nu_e$'s in a $\nu_\mu$ beam with the appearance of $\overline{\nu}_e$'s in a $\overline{\nu}_\mu$ beam in its second three year run, given sufficient funding NO$\nu$A may be able to separate the octant and $\theta_{13}$, which determine the total $\nu_e+\overline{\nu}_e$ appearance, from the hierarchy and $\delta$, which roughly determine the difference $\nu_e-\overline{\nu}_e$.  However for most values of $\delta$ even the combination of T2K and NO$\nu$A will not be able to break the degeneracy between $\delta$ and the hierarchy, and so will not be able to determine either.

Fortunately the large value of  $\theta_{13}$ also implies that 1-3 neutrino oscillations, those with amplitudes proportional to $\spp2213$, 
are large enough to be observed in reactor neutrino experiments at medium baselines (40-60 km).  These oscillations are almost periodic in the inverse energy $E$, but their small aperiodicity may be used to determine the neutrino mass hierarchy \cite{petcovidea} thus breaking the degeneracy at T2K and NO$\nu$A and so allowing a determination of the CP-violating Dirac phase.  More precisely, combined with a hierarchy determination, the NO$\nu$A appearance mode can only determine sin($\delta)$ because the transformation $\delta\rightarrow\pi-\delta$ is degenerate with a slight shift in $\theta_{23}$ or $\theta_{13}$.  However the effective mass difference determined by a reactor experiment, which will be discussed below, differs from the $\delta$-dependent atmospheric effective mass \cite{parke2005} determined by the disappearance channel of experiments such as MINOS and NO$\nu$A.  This difference may yield a 1$\sigma$ determination of cos($\delta$).  Such reactor experiments are not only possible but indeed {\it will}\ be performed within the next decade~\cite{renonuturn,yifangseminario}.  

Furthermore, these reactor neutrino experiments may well have two detectors so as to reduce their sensitivity to systematic errors resulting from the detector's unknown energy response.  This would offer a unique opportunity to measure $\delta$ using a strategy similar to that of the DAE$\delta$ALUS experiment described in Ref.~\cite{dead}.  One would need a high intensity stationary pion source located at about 10 km from one detector and 20 km from the other.  The fact that only one source is required makes such an experiment both cheaper than DAE$\delta$ALUS would be at LBNE and also more precise, as it eliminates the uncertainties due to the relative strengths of the sources.

In this letter we present the main results of a series of simulations of medium baseline reactor experiments.    We will present the reliability of the determinations of the hierarchy in such experiments and will also find the optimal baselines for their detectors.  

\section{Simulations}

We consider reactor experiments in which $\overline{\nu}_e$'s are detected via inverse $\beta$ decay by a detector with a 20 kton target mass consisting of 12\% hydrogen.  This is the mass of the proposed Daya Bay II detector, but it is 1.1 times the mass of the proposed detector for RENO 50, and so to interpret the results below for RENO 50  one needs to multiply all times by a factor of 1.1.  We assume a detector resolution of $3\%/\sqrt{E\rm{(MeV)}}$, where $E$ is the prompt energy which is $0.8$ MeV less than the $\overline{\nu}_e$ energy.  We assume a perfectly understood detector energy response and also ignore backgrounds except for the simulations reported in the last section of this note, which include a simplified model of geoneutrinos and include various models of unknown energy response.  As a result of these approximations our results are overly optimistic. 

We have performed 3 and 6 year simulations using several reactor flux models with various cutoffs using fluxes arising from over 100 combinations of baselines.  For each combination, we have simulated 5,000 experimental runs with each hierarchy.  The data analysis methods described below are then used to attempt to determine the hierarchy.  We report the percentage $p$ of experiments for which the hierarchy is determined correctly.

In most of our simulations we use the value of $\m21$ from Ref.~\cite{pdb}, $\spp2212$ from Ref.~\cite{gando}, $\mn32$ determined by combining $\nu$ and $\overline{\nu}$ mass differences from Ref.~\cite{minosneut2012}  and $\spp2213$ from \cite{neut2012} 
\bea
&&\m21=7.59\times 10^{-5}{\mathrm{\ eV^2}}\hsp
\spp2212=0.857\nonumber\\
&&\mn32=2.41\times 10^{-3}{\mathrm{\ eV^2}}\hsp
\spp2213=0.089.\nonumber
\eea
We have also systematically studied the effects of shifting these parameters, as will be described below.  The fact that we hold these parameters fixed affects the chance of successfully determining the hierarchy.  In the last section of our note we will report the results of another series of simulations in which we allowed the most relevant of these parameters to vary according to their current experimental uncertainties.

The determination of the neutrino mass hierarchy from 1-3 oscillations in the $\overline{\nu}_e$ spectrum  proceeds as follows.  At energies $E$/MeV greater than $L$/12 km, deviations from periodicity in $1/E$ are too small to be measured and the wavenumber $k$ determines \cite{parke2005}
\beq
\meff=\cp212\mn31+\sp212\mn32. \label{meff}
\eeq
At low energies the deviation from periodicity is large and it determines various combinations of $\mn31$ and $\mn32$ given in Ref.~\cite{noiteor}, for example the energy of the 16th oscillation peak is proportional to $\mn31$.   To determine the mass hierarchy one needs to combine two distinct combinations of $\mn31$ and $\mn32$, thus one must combine the high and low energy parts of the spectrum.   For example, the mass hierarchy is normal (inverted) if $\mn31$ is greater (less) than $\meff$. 

The most studied algorithm which determines the hierarchy given a reactor $\overline{\nu}_e$ spectrum is that of Ref.~\cite{caojun}.  One first finds the Fourier transform of the measured spectrum, as suggested in Ref.~\cite{hawaii}.  1-3 oscillations have a wavenumber $k$ of about $\mn32$ and so the peak structure of the transform at $k\sim\mn32$ is sensitive to these oscillations and thus the hierarchy.  In Refs.  \cite{caojun, caojun2} it was shown that the heights of the peaks can be combined into two real numbers $RL$ and $PV$ such that $RL+PV$ is positive if and only if the hierarchy is normal.  In Ref.~\cite{noiteor} two more observables were added, one mixing information from the Fourier sine and cosine transforms and one using a nonlinear Fourier transform with the same aperiodicity as the 1-3 oscillations.  We will report analyses of our simulations using both $RL+PV$ and also using a neural network which finds the combination of all four observables which best determines the hierarchy.

\section{Challenges}

\begin{figure} %[!tph]
\begin{center}
\includegraphics[width=3.1in,height=1.35in]{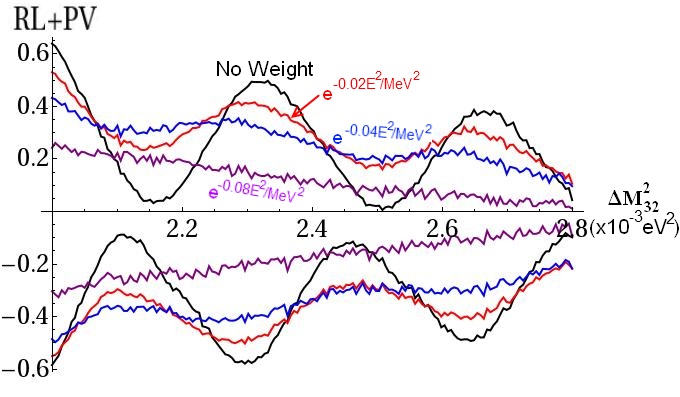}
\caption{Simulated average values of $RL+PV$ obtained from 100,000 neutrinos observed at a baseline of 58 km assuming the numerically interpolated reactor spectra from the 1980's \cite{uflusso,pflusso,quadflusso}.  The black curve uses an unweighted Fourier transform while the purple, blue and red curves respectively use weights of  exp$(-0.02E^2/\rm{MeV}^2)$, exp$(-0.04E^2/\rm{MeV}^2)$ and exp$(-0.08E^2/\rm{MeV}^2)$.  Notice that the first weight is so flat that the oscillations remain quite large, while the last is so strong that the $RL+PV$ signal is suppressed.}
\label{mediafig}
\end{center}
\end{figure}

An obstruction to this analysis has been described in Ref.~\cite{oggi}, which observed that $RL+PV$ is very sensitive to the choice of model of the reactor neutrino flux and to variations of $\meff$  smaller than the precision of its determination by MINOS \cite{minos}.   Following Ref.~\cite{noispurioso} we have reduced this spurious dependence,
%To eliminate the spurious dependence upon the shape of the tail, 
by employing a weighted Fourier transform in which higher energies are weighted less heavily, providing a soft cutoff.  Fig.~\ref{mediafig} shows the average values of $RL+PV$ in simulated data using a normal and a weighted Fourier transform.  The normal Fourier transform leads to the fluctuating dependence found in Ref.~\cite{oggi}, but these fluctuations are much smaller with a weighted transform.  A steeper weight would further reduce the amplitude of the fluctuations, but  would also reduce the average value of $RL+PV$, weakening the hierarchy dependence.   Thus we find that the spurious dependence observed in Ref.~\cite{oggi} can be eliminated via a simple modification of the analysis.

\begin{figure} %[!tph]
\begin{center}
\includegraphics[width=2.8in,height=1.62in]{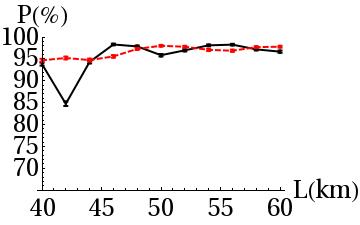}
\caption{The probability of success, calculated using $RL+PV$, with simulations of 6 years of flux arising from a reactor complex at a distance $L$ and another complex 500 meters further away.  The error bars are the statistical errors arising from the number of simulations.  The black solid and red dashed curves correspond to simulations using the quadratic fit to the reactor flux models of the 1980's \cite{quadflusso} and the quintic fit to the recent calculation of Ref.~\cite{huber}.}
\label{mfig}
\end{center}
\end{figure}

\begin{figure} %[!tph]
\begin{center}
\includegraphics[width=2.8in,height=1.62in]{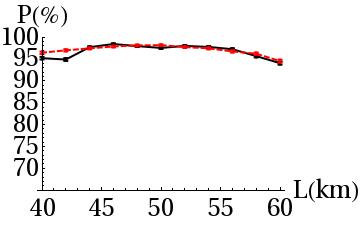}
\includegraphics[width=2.8in,height=1.62in]{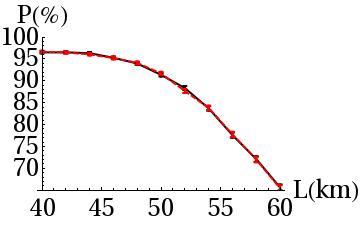}
\caption{As in Fig.~\ref{mfig}, but now including a shallow weight of  $e^{-0.02E^2}$ in the upper panel and a steeper weight of  $ e^{-0.08E^2}$ in the lower panel.  Note that in both cases the model dependence of Fig.~\ref{mfig} reduced, suggesting that this dependence is caused by neutrinos in the high energy tail.  The later spectrum reduces the dependence by more, but at a cost of hierarchy-dependent information at moderately high energies and as a result increases statistical errors arising from the limited number of events in each bin.  As a result the first weight outperforms the second at the long baselines where statistical errors are important.}
\label{mwfig}
\end{center}
\end{figure}

Does our choice of flux model matter?  Different flux models differ primarily in the high energy tail.  Therefore the choice of flux models does have a large effect on the unweighted RL+PV, as can be seen in Fig.~\ref{mfig}.  However, once weights are included, the choice of flux model at most baselines is important at about the same level as the statistical errors arising from our finite number of experiments, as can be seen in Fig.~\ref{mwfig}.  Therefore we will choose the quadratic flux model of  Ref.~\cite{quadflusso} for easier comparison with previous studies, such as that of Ref.~\cite{caojun2}.

 We will now discuss a more serious challenge which afflicts otherwise promising sites for such an experiment.%, such as BaiMianShi to the north east of Daya Bay \cite{noisim}. 

As a determination of the hierarchy requires a medium baseline, the neutrino flux arriving from each reactor will necessarily be quite low.  This means both that the detector must be very large and also that flux from multiple reactors must be used.  Multiple reactors are available, especially in places like Japan, Korea and China's Guangdong province where such experiments may be built.  However multiple reactors imply multiple baselines, and so $\overline{\nu}$s will arrive at the detector with their 1-3 oscillations out of phase.  Neutrinos from different reactors are not coherent with each other, wavefunctions are not added, but probabilities are added and this destroys the fine structure of the spectrum whose precise measurement is essential to a determination of the hierarchy.

\section{Results}

We now provide the most systematic analysis of this interference effect to date.  Our goal is to illustrate the effect of multiple baselines on the chance of success of the experiment and on the optimal location.  We will consider two 18 GW thermal capacity idealized point reactor neutrino sources.   By defining an effective baseline difference, our results can be applied to configurations with many reactors. The reactor flux is normalized such that, including $\nu$ oscillations, at 58 km each 18 GW complex yields 25,000 $\overline{\nu}_e$  in 3 years.

\begin{figure} %[!tph]
\begin{center}
\includegraphics[width=2.8in,height=1.62in]{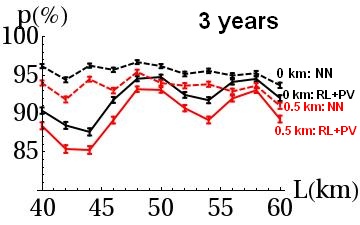}
\includegraphics[width=2.8in,height=1.62in]{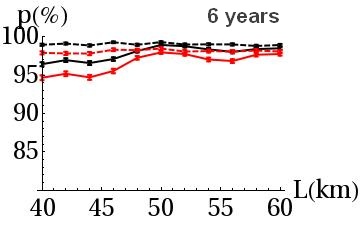}
\caption{Chance of success, $p$, after 3 years (first panel) and 6 years (second panel) of live time, with neutrinos from two 18 GW reactor complexes at distinct baselines.  The horizontal axis is the shortest baseline, the color is the difference between baselines: 0 (black) and 500 m (red). The solid curves use $RL+PV$ and the dashed curves a neural network.}
\label{int01fig}
\end{center}
\end{figure}

\begin{figure} %[!tph]
\begin{center}
\includegraphics[width=2.8in,height=1.62in]{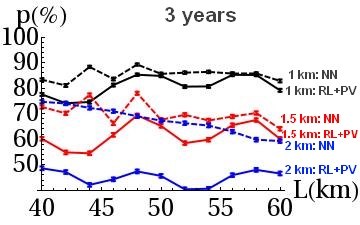}
\includegraphics[width=2.8in,height=1.62in]{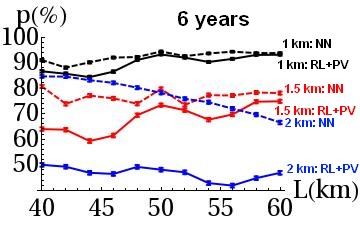}
\caption{As in Fig. \ref{int01fig} but now the baseline differences are 1 km (black), 1.5 km (red) and 2 km (blue).}
\label{int12fig}
\end{center}
\end{figure}

\begin{figure} %[!tph]
\begin{center}
\includegraphics[width=2.8in,height=1.62in]{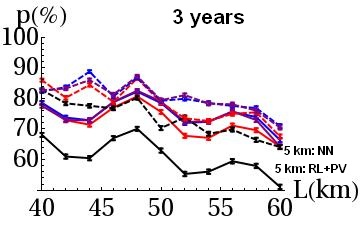}
\includegraphics[width=2.8in,height=1.62in]{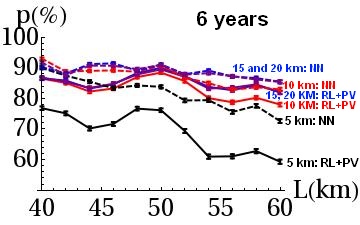}
\caption{As in Fig. \ref{int01fig} but now the baseline differences are  5 km (black), 10 km (red), 15 km (blue) and 20 km (purple).}
\label{int520fig}
\end{center}
\end{figure}

In Figs.~\ref{int01fig},\ \ref{int12fig} and \ref{int520fig} we display the probability $p$ of successfully determining the hierarchy with 3 and 6 years of live time for various combinations of baselines.  The solid and dashed curves are analyses using respectively $RL+PV$ and a neural network optimizing 24 coefficients corresponding to the 4 hierarchy indicators of Ref.~\cite{noiteor} and to 6 weights: $e^{-0.08E^2}\pic$\ $ e^{-0.02E^2}\pic$\ $ e^{-0.04(E - 3.6)^2}\pic$\ $ e^{-E/8}\pic$\ $ e^{-0.1(E - 5.25)^2}$\ and\ $e^{-(E - 3.6)^3/100}$.   In all, the weighted cosine and sine Fourier transforms used for the $i$th indicator are
\bea
F^i_c(k)&=&\sum_j w^i(E_j) N(E_j)  \cos\left(\frac{kL}{E_j}\right)\\
F^i_s(k)&=&\sum_j w^i(E_j) N(E_j)  \sin\left(\frac{kL}{E_j}\right) \nonumber\\
w^i(E)&=&a_1^{\ i} e^{-0.08\left(\frac{E}{\rm{MeV}}\right)^2}+a_2^{\ i}e^{-0.02\left(\frac{E}{\rm{MeV}}\right)^2}\nonumber\\&&+\hspace{-.0cm}a_3^{\ i}e^{-0.04\left(\frac{E}{\rm{MeV}}-3.6\right)^2}\hspace{-0cm}+a_4^{\ i}e^{-\frac{E}{8\ \rm{MeV}}}\hspace{-0cm}\nonumber\\
&&+\hspace{0cm}a_5^{\ i}e^{-0.1\left(\frac{E}{\rm{MeV}}-5.25\right)^2}\hspace{-0.1cm}+\hspace{-0.05cm}a_6^{\ i}e^{-0.01\left(\frac{E}{\rm{MeV}}-3.6\right)^3}.\nonumber\label{fcos}
\eea
As can be seen in Fig.~\ref{mediafig}, if the weights suppress the high energy tail too weakly, such as $e^{-0.02\left(\frac{E}{\rm{MeV}}\right)^2}$, then the spurious fluctuations persist.  On the other hand if they suppress it too strongly, such as $e^{-0.08\left(\frac{E}{\rm{MeV}}\right)^2}$, then statistical errors are increased and so the $RL+PV$ hierarchy signal is weakened.  The weights used here have been chosen to strike a balance between these two effects.

To determine the coefficient matrix $a$, each run of 5,000 simulations/hierarchies was divided into groups.  For each group, we determined the chance of success independently by using coefficients obtained by training the neural network on different groups.  In order to avoid overfitting, coefficients obtained by training the neural network on a given group are never used to evaluate its chance of success.  The probability of success that is reported below is the average probability of success of all of the groups.  Note in particular that the optimal coefficients are distinct for each experimental run, for example the nonlinear Fourier transform is much more heavily weighted at baselines below about 54 km, where it significantly outperforms the older indicators.

\begin{figure} %[!tph]
\begin{center}
\includegraphics[width=2.8in,height=1.62in]{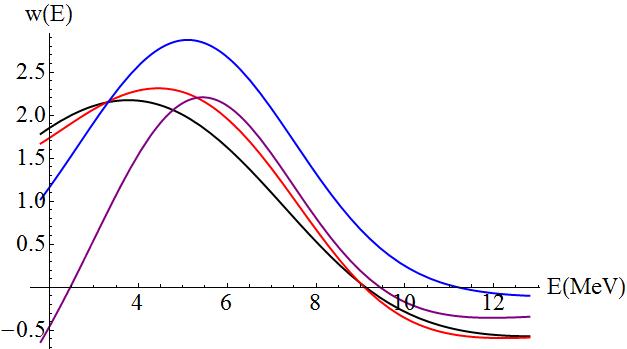}
\caption{The weight functions $w^i(E)$ for $RL$ (black), $PV$ (red), the mixed sine cosine indicator of Ref.~\cite{noiteor} (blue) and the weighted Fourier transform of Ref.~\cite{noiteor} (purple) which optimize the chance of successfully determining the neutrino mass hierarchy in 6 years with one reactor complex at 52 km and another at 52.5 km}
\label{wfig}
\end{center}
\end{figure}

For example, in Fig.~\ref{wfig} we illustrate the 4 optimal weights $w^i(E)$ in the case most relevant to Daya Bay II, corresponding to 6 years of exposure to one reactor complex at 52 km and another at 52.5 km.  The corresponding coefficient matrix is
\beq
a\hspace{-.08cm}=\hspace{-.08cm}\left(
\begin{array}{cccccc}
0.07&2.53&1.71&-3.58&0.19&0.61\\
-0.35&0.93&1.73&-3.35&-0.13&2.13\\
-1.61&0.91&1.83&-1.51&3.71&0.86\\
0.05&0.66&-0.48&-1.79&3.66&-0.51\\
\end{array}
\right).
\eeq
With the number of simulations that we have run, these optimal coefficient matrices are fairly stable against statistical fluctuations.  However they do exhibit some dependence upon the baselines and the running time of the experiment.

Figs.~\ref{int01fig}\ and \ref{int12fig} present the 3 and 6 year probabilities of success for baseline differences between 0 and 2 km.  A 2 to 5 km baseline difference causes a lower chance of success.  With these analysis methods, a 95\% chance of a successful hierarchy determination in 6 years is only possible if the baseline difference is appreciably less than 1 km, about the effective difference for the BaiMianShi and Mudeungsan sites \cite{noisim} for Daya Bay II and RENO 50.  As the preferred Guemseong site for RENO 50 receives neutrinos from only 18 GW of reactors,  the times reported here must be doubled, in addition to the factor of 1.1 discussed earlier.   This is not the case for the Munmyeong or Munsusan sites.

Fig. \ref{int520fig} illustrates that, for baseline differences of 5 to 20 km, the far detector is a background.  As a result a larger baseline difference and a shorter baseline to the near reactor both increase the signal/background ratio and so $p$.   These strong backgrounds reduce the optimal baseline.

Interference and flux from more distant reactors is a big problem for differences above  500 meters, arising when a detector is not perpendicular to a long  linear array like the Daya Bay - Ling Ao complex. Sites such as BaiYunZhang are perpendicular and so enjoy identical baselines, but use flux from only a single array.  In particular Daya Bay II sites such as BaiMianShi are not competitive.  While the old sites for Reno 50 will face similar problems, as the largest mountains are at a 45 degree angle to the reactor array.  New sites such as Guemseong and Munmyeong have negligible interference, although the latter must endure reasonably large backgrounds from reactors 130 km away.  Munsusan on the other hand will suffer from a baseline difference of order 700 meters.

The minimal baseline difference at potential Daya Bay II sites that use flux from multiple reactor complexes is about 500 meters, corresponding to the Dongkeng site of Ref.~\cite{noisim} which uses flux from the Taishan and Yangjiang complexes.  Thus we find that the best case probability of determining the hierarchy is about 98\%.  This result depends upon the values of the neutrino mixing parameters, in general we have found that a 1$\sigma$ increase in $\sp213$ or $\mn21$ or a 1$\sigma$ decrease in $\mn32$ can improve the hierarchy determination by $0.1\sigma-0.3\sigma$.  

The {\it{disappearance}} channel at NO$\nu$A may provide a $1-2\%$ determination of the atmospheric effective mass of Refs.~\cite{okamura,parke2005}.  For $\delta\sim\pi$, this mass differs from the high energy reactor effective mass (\ref{meff}) by about $1.5\%$ and from the low energy reactor effective mass $\mn31$ by nearly $3\%$.  As the sign of these differences depends upon the hierarchy, NO$\nu$A disappearance data can improve the hierarchy determination at Daya Bay II.  However there is no such advantage if $\delta\sim 0$, as the atmospheric effective mass would be nearly equal to that of Eq.~(\ref{meff}), which will measured more precisely at Daya Bay II.

\section{Sources of Error} 

In this note we have determined the probability that, using various Fourier transform based analyses, a medium baseline reactor neutrino experiment can successfully determine the hierarchy.   This chance of success and the associated number of $\sigma$'s is a somewhat pessimistic indicator of the sensitivity of the experiment because those experiments that fail typically have low values of $|RL+PV|$ due to statistical fluctuations.  As we have not considered systematic errors, this situation can be cured by simply running the experiment longer.  More to the point, the median experiment, defined as an experiment yielding the {\it{median}} value of $|RL+PV|$, provides the correct hierarchy with significantly more confidence than the {\it{mean}} chance of success reported here.  The results in Ref.~\cite{statistica} suggest for example that a 98\% mean chance of success corresponds to a 3.5$\sigma$ confidence for the median experiment.  While that study used a $\chi^2$ analysis and not a Fourier analysis, whenever it was possible to cross check the two approaches we have found mutually consistent results.  A similar consistency was found in Ref.~\cite{oggi}.

At the same time, the results of this study are in many ways overoptimistic.  A number of sources of error have been ignored which are certain to degrade the sensitivity to the hierarchy.  The most prominent among these is the unknown nonlinear energy response of the detector.  The energy of a reactor neutrino is determined by the number of photoelectrons measured by the various photomultiplier tubes (PMTs).  As Daya Bay II and RENO 50 will have detectors whose diameters are of order the mean free path of light inside of their liquid scinitillators, the number of photoelectrons also strongly depends on the position of the event.  The position will be determined roughly by the distribution of photoelectrons among the PMTs and also by the timing at which the photons are detected in various locations.  

However a determination of the number of photoelectrons expected for each position of the inverse $\beta$ decay event is very challenging.  It will require an extensive calibration campaign, and even then it is unclear whether it will reach the subpercent precision required for a determination of the hierarchy.  In addition it is complicated by other factors, such as energy lost by the geometry-dependent Cherenkov radiation from the positrons and dead time in the electronics, as well as the instability of the scintillator itself over the long timescales over which such experiments will run.   

The unknown energy response of the detector may well prove to be too great of an obstacle for the hierarchy to be determined.  However, even if the target $1\%$ knowledge of the energy response is achieved, it will still introduce a large systematic error into the spectrum which will reduce the confidence in a hierarchy determination.  This error can be significantly reduced in a setup with 2 identical detectors \cite{noi6,snowmass,noi6sim}.  However it not only significantly impacts the chance of successfully determining the hierarchy, but also the magnitude of its effect depends on its unknown energy and position dependence.  Below we will consider its affect on the chance of success in the case of a simple model of the detectors energy response.

Much of the hierarchy determining power \cite{yifang,statistica} of these experiments will come from a comparison of $\meff$\ with the atmospheric splitting \cite{parke2005} determined from the disappearance channel at accelerator experiments like NO$\nu$A.   This method relies upon a 1\% difference between between the effective splittings whose sign determines the hierarchy.  However the atmospheric splitting itself also suffers from systematic effects, for example the nuclear effects of Ref.~\cite{huber2013} can increase the measured atmospheric splitting by 2\%, enough for the normal hierarchy to mimic the inverted hierarchy.

The good news is that some of the flux uncertainties which are important for the determination of $\theta_{13}$ and the reactor anomaly \cite{reattoreanom,noiunokm} are irrelevant for the determination of the hierarchy.  The determination of the hierarchy relies entirely upon the locations of small, short oscillations in the spectrum and not upon the broad features caused by the uncertainties in flux models and energy dependent acceptance.  This insensitivity is built into the Fourier approach, as broad features affect the Fourier transform only at low wave numbers, far below those where the peaks and valleys determining $RL$ and $PV$ are located. Indeed, this is one of the main motivations for the Fourier approach.

The main remaining sources of error are the uncertainties in the neutrino mass matrix and various backgrounds.  Both of these are capable of affecting the small scale structure of the spectrum and so the determination of the hierarchy.  The mass matrix elements whose uncertainties have the largest affect on the determination of the hierarchy are $\theta_{13}$ and $\mn32$.  There are several relevant backgrounds, the most important of which are of comparable size for these experiments.  We will consider just one, a simplified model of geoneutrinos employed just to see what effect it has.  %We will see that its effect is well within the statistical errors of our simulations, whereas the errors caused by $\theta_{13}$ and $\mn32$ are statistically significant.

\begin{figure} %[!tph]
\begin{center}
\includegraphics[width=3.1in,height=1.55in]{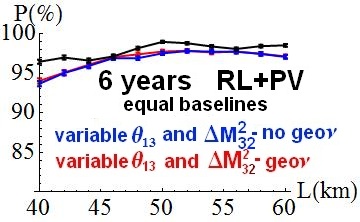}
\includegraphics[width=3.1in,height=1.55in]{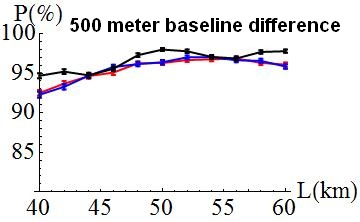}
\includegraphics[width=3.1in,height=1.55in]{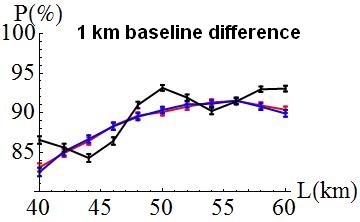}
\includegraphics[width=3.1in,height=1.55in]{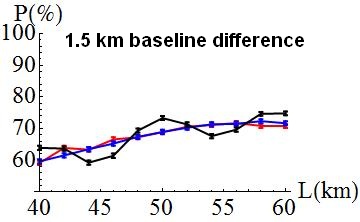}
\includegraphics[width=3.1in,height=1.55in]{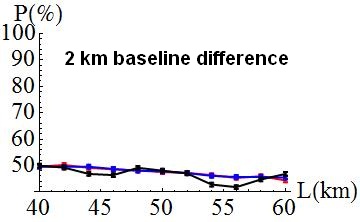}
\caption{The chance of success in 6 years as determined using $RL+PV$.  The neutrinos arise from two sources whose baseline differences vary from 0 to 2 km.  The black curves are the results of Figs.~\ref{int01fig} and \ref{int12fig}.  In the blue and red curves the parameters $\theta_{13}$ and $\mn32$ are Gaussian distributed with errors corresponding to the uncertainties to which they have been determined experimentally.  The red curve also contains a simple model of a geoneutrino background.}
\label{rlfig}
\end{center}
\end{figure}

\begin{figure} %[!tph]
\begin{center}
\includegraphics[width=3.1in,height=1.55in]{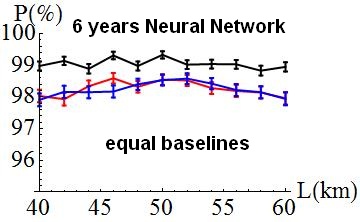}
\includegraphics[width=3.1in,height=1.55in]{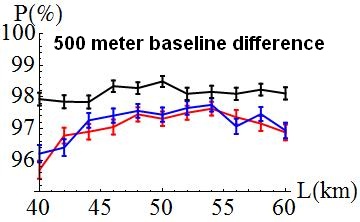}
\includegraphics[width=3.1in,height=1.55in]{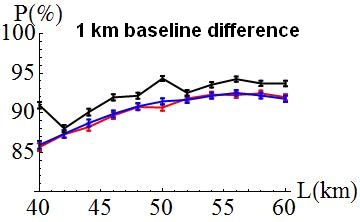}
\includegraphics[width=3.1in,height=1.55in]{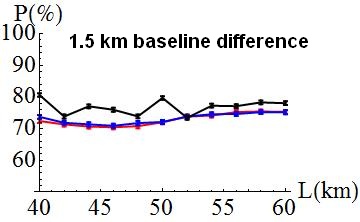}
\includegraphics[width=3.1in,height=1.55in]{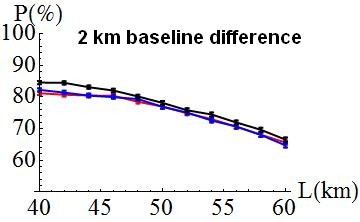}
\caption{As in Fig.~\ref{rlfig} but the analysis is done using a neural network.}
\label{nnfig}
\end{center}
\end{figure}

For the analysis of errors we consider 6 year runs with 2 neutrino sources with baselines which differ by 0, 500 meters, 1 kilometer, 1.5 kilometers and 2 kilometers.  The results of our study, displayed in Figs.~\ref{rlfig} and \ref{nnfig} for the $RL+PV$ and neutral network analyses respectively, are compared with the results of Figs.~\ref{int01fig} and \ref{int12fig} which correspond to the black curves in those figures.  In each new simulation $\spp2213$ and $\mn32$ are chosen from a Gaussian distribution centered at $0.089$ and $2.4\times 10^{-3}$\ eV respectively.  These Gaussians have width $\sigma$ equal to $0.013$ and $10^{-4}$\ eV.  The blue curves correspond to simulations in which no backgrounds are included.  The red curves also include a simple model of geoneutrinos.  The total number of detected geoneutrinos in each simulation follows a Gaussian distribution centered at 2520 with a width or $20\%$.  The energy distribution of the geoneutrinos is centered at $1.8$ MeV with a width of 500 keV.   The finite resolution of the detector is not convolved with the geoneutrino energy distribution, instead the width is taken to model the sum of the true width and the effect of finite resolution.

The error bars shown reflect only statistical errors in the simulations, caused by the fact that for each setup we only ran 5,000 simulations with each hierarchy.  The blue and red curves agree to within these error bars, and so geoneutrinos contribute no noticeable effect to within the precision of our simulations.  On the other hand, the effect of the uncertainty in $\theta_{13}$ and $\mn32$ is significant.   First of all, the fluctuating baseline dependence observed in Figs.~\ref{int01fig} and \ref{int12fig} and then reported as the black curve in Figs.~\ref{rlfig} and \ref{nnfig} is eliminated once uncertainties in the oscillation parameters is considered.  In the case of the $RL+PV$ analysis in Fig.~\ref{rlfig}, the effect of this parameter variation on the central values of the probability of success is quite small.  Considering the uncertainty in $\theta_{13}$ and $\mn32$ one finds that the chance of success is reduced by of order $1\%$ when the baseline difference is 0 or 500 meters, but is effectively unchanged when the difference is 2 km.

On the other hand, for all baseline differences one can see in Fig.~\ref{nnfig} that the parameter uncertainty reduces the effectiveness of the neural network, although it continues to enjoy a significantly higher chance of success than $RL+PV$.  This is because if the neutrino mass matrix is well known, the neural network optimizes itself to the corresponding parameters.  

Once Daya Bay has completed its 3 year run, $\theta_{13}$ will be measured more precisely than has been considered here.  Similarly, before Daya Bay II and RENO 50 begin, NO$\nu$A and MINOS+ will have measured $\mn32$ more precisely than has been considered here.  These more precise values can be used to train the neural network, and so in this sense the final effect of the uncertainty of $\theta_{13}$ and $\mn32$ on the chance of success will be smaller than the $0.1-0.2\sigma$ indicated in Fig.~\ref{nnfig}.  Nonetheless, there are other albeit perhaps smaller sources of error which have not been considered.

There are several potentially catastrophic sources of error at such experiments.  For example, if after some time the scintillator's optical properties change as occurred at the Palo Verde or Chooz experiments, due to the large size of the detector the number of photoelectrons would drop so as to make the resolution too poor to observe 1-3 oscillations.  As a result a determination of the hierarchy would be impossible.  Even a change in optical properties similar to that observed in the first year of running at Daya Bay may have serious consequences for the hierarchy determination.  

Another concern is the coherence of the first and second neutrino mass eigenstates with respect to the third eigenstate.  At a fixed energy $E$, during a 50 km voyage from the reactor to the detector,  their relative velocities will lead to a relative separation of  roughly 
\beq
\Delta L=\frac{\mn32}{2 E^2}(50{\mathrm{\ km}})=\frac{6\times 10^{-11}m}{(E/{\rm{MeV}})^2}.
\eeq
If $\Delta L$ is of order the length of the wave packet, the 1-3 oscillation amplitude will be reduced and if the wave packets are shorter then there will be no 1-3 oscillations \cite{boris,smirnov}.   

This threshold is an order of magnitude larger than the lower bound on the coherence length that may be obtained from the observation of 1-2 oscillations at KamLAND \cite{gando}, which is marginally stronger than bounds that may be obtained from the observation of 1-3 oscillations at Daya Bay \cite{neut2012} and RENO \cite{reno}.  Therefore it appears as though current experimental bounds do not preclude a coherence length of the neutrino wave packets shorter than $\Delta L$, leaving open the possibility that decoherence of the neutrino mass eigenstates will be observed at medium baseline reactor experiments and as a result the hierarchy will not be determined.  Curiously, a coherence length of order $10^{-12}$m MeV${}^2/E^2$  would imply that partial decoherence has already reduced the $1-3$ oscillation amplitude at Daya Bay and to a lesser extent at RENO.  As a result the true value of $\theta_{13}$ would be increased, reducing the tension with the recent accelerator determination of $\theta_{13}$ by T2K \cite{t2k}.  On the other hand the coherence length is unlikely to be more than an order of magnitude longer than $\Delta L$ as a result of the phase shift caused by the nucleon recoil energy in the radioactive decay in which these neutrinos are created.

\begin{figure} %[!tph]
\begin{center}
\includegraphics[width=3.1in,height=1.55in]{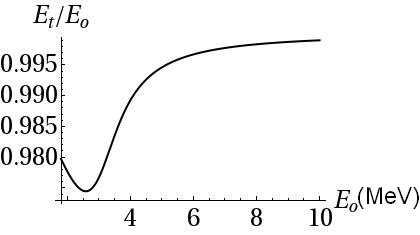}
\caption{The ratio of the observed to true energy corresponding to the case $\epsilon=1$ of the unknown energy response model~(\ref{xin}).  This model corresponds to an unknown nonlinear energy response of about 2\%.}
\label{xinmodelfig}
\end{center}
\end{figure}

Of the sources of error which are certain to be present, perhaps the most serious is the detector's unknown energy response. Define $E_{\rm{o}}$ to be the energy which would on average be deduced for a neutrino of energy $E_{\rm{t}}$ using the results of all of the calibrations and simulations available.  In other words $E_{\rm{o}}$ is the best guess that the experimenter will be able to make for the energy of a given neutrino when analyzing the data, ignoring statistical fluctuations.  Below we have considered a simple model of this response
\bea
E_{\rm{o}}&=&\frac{2\mn32{\rm{(eV^2)}}+4\epsilon\cp212\m21{\rm{(eV^2)}}{L{\rm(m)}}}{2\mn32{\rm{(eV^2)}}+\frac{\phi}{1.27}\frac{E_{\rm{t}}{\rm{(MeV)}}}{L{\rm(m)}}}E_{\rm{t}}\nonumber\\
&&+\frac{(1-2\epsilon)\frac{\phi}{1.27}\frac{E_{\rm{t}}{\rm{(MeV)}}}{L{\rm(m)}}}{2\mn32{\rm{(eV^2)}}+\frac{\phi}{1.27}\frac{E_{\rm{}}{\rm{(MeV)}}}{L{\rm(m)}}}E_{\rm{t}} \label{xin}
\eea
where
\beq
{\rm{sin}}(\phi)=\frac{\cp212{\rm{sin}}\left(\frac{2.54\m21{\rm{(eV^2)}} L{\rm{(m)}}}{E_{\rm{t}}{\rm{(MeV)}}}\right)}{\sqrt{1-\spp2212{\rm{sin}}^2\left(\frac{2.54\m21{\rm{(eV^2)}} L{\rm{(m)}}}{E_{\rm{t}}{\rm{(MeV)}}}\right)}} .
\eeq
Here $\epsilon$ is a parameter such that $\epsilon=0$ corresponds to a perfectly understood energy response and $\epsilon=1$ corresponds to the model which was proposed in Ref.~\cite{oggi} as the worst case for the determination of the hierarchy.  More precisely, the case $\epsilon=1$ corresponds to a systematic error in the energy response which causes the normal hierarchy spectrum to resemble the inverted hierarchy spectrum.  The ratio of the true to the observed energy in the case $\epsilon=1$ is drawn in Fig.~\ref{xinmodelfig}.

\begin{figure} %[!tph]
\begin{center}
\includegraphics[width=3.1in,height=1.55in]{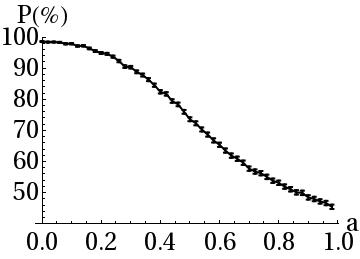}
\caption{The probability of correctly determining the hierarchy with 120 kton years of exposure to 36 GW of reactors at 54 km given the unknown energy response model of Eq.~(\ref{xin}) for various values of the parameter $\epsilon$.}
\label{xinfig}
\end{center}
\end{figure}

We consider 120 kton years of exposure to 36 GW of thermal capacity of reactors at a single baseline of 54 km, using an RL+PV analysis.   While the nonlinear response considered here poses a more serious problem in the case of the normal hierarchy than the inverted hierarchy, there is a similar model which is more dangerous in the case of the inverted hierarchy.  Therefore we will report the average chance of success, corresponding to the average of the chance of correctly determining the hierarchy in the case in which the true hierarchy is normal and the chance in the case in which the true hierarchy is inverted.  The results of 5,000 simulations with each hierarchy and 50 values of $\epsilon$ are reported in Fig.~\ref{xinfig}.

One may observe that for $\epsilon$ greater than about $1/3$, corresponding to an unknown energy response of about $0.7\%$, the hierarchy can only be determined correctly in $89\%$ of simulations using the $RL+PV$ Fourier transform based analysis.  As has been anticipated in Ref.~\cite{noiteor}, this partially reflects a weakness in the Fourier analysis.  Indeed, in the absence of nonlinearity the $\chi^2$ approach of Ref.~\cite{noi6sim} obtained  $\Delta\chi^2=18$ which according to Eq.~(9) of Ref.~\cite{statistica} corresponds to a probability of successfully determining the hierarchy of 98.8\%, similar to that obtained using the Fourier transform approach with a neural network.  In the case $\epsilon=1/3$ the $\chi^2$ approach of Ref.~\cite{noi6sim} yielded $\Delta\chi^2=12$ corresponding to a probability of success of 97.6\%, appreciably better than the probability of success obtained here using the Fourier transform.  

However, the apparent superiority of the $\chi^2$ approach in the presence of an unknown energy response results from the use of pull parameters to parametrize this response.  In Ref.~\cite{noi6sim} three such parameters were introduced and chosen so as to minimize $\chi^2$.  We have repeated this analysis with no pull parameters and obtained $\Delta\chi^2=2$ for the case $\epsilon=1/3$, which is much worse than the confidence obtained above using the Fourier transform analysis.  The reason for this is that the Fourier analysis is insensitive to the overall mass scale, and so is only adversely affected by a part of the unknown energy response.   One may apply an analogous pull parameter procedure to a Fourier method, to obtain a probability of success of order 97\%.  However by using a particular functional form of the spectrum to determine the pull parameters it would sacrifice much of the model independence which is the strength of the Fourier approach.

\section* {Acknowledgement}

\noindent
We would like to thank C. Bonati for useful discussions and suggestions.  JE is supported by the Chinese Academy of Sciences
Fellowship for Young International Scientists grant number
2010Y2JA01. EC and XZ are supported in part by the NSF of
China.  

%%%%%%%%%%%%%%%%%%%%%%%%%%%%%%%%

\end{document}